\documentclass[sigconf, screen]{acmart}
\usepackage{balance}

\usepackage{algorithm}
\usepackage[noend]{algorithmic}
\usepackage{multirow}
\usepackage{booktabs} 
\usepackage{threeparttable}
\usepackage[shortlabels]{enumitem}

\acmPrice{15.00}
\acmDOI{10.1145/3236024.3264835}
\acmYear{2018}
\copyrightyear{2018}
\acmISBN{978-1-4503-5573-5/18/11}
\acmConference[ESEC/FSE '18]{Proceedings of the 26th ACM Joint European Software Engineering Conference and Symposium on the Foundations of Software Engineering}{November 4--9, 2018}{Lake Buena Vista, FL, USA}
\acmBooktitle{Proceedings of the 26th ACM Joint European Software Engineering Conference and Symposium on the Foundations of Software Engineering (ESEC/FSE '18), November 4--9, 2018, Lake Buena Vista, FL, USA}

\begin{document}
\title{DLFuzz: Differential Fuzzing Testing of Deep Learning Systems}

\author{Jianmin Guo}
\affiliation{%
  \institution{KLISS, BNRist, School of Software, Tsinghua University, China}
}
\email{guojm17@mails.tsinghua.edu.cn}

\author{Yu Jiang}
\affiliation{%
  \institution{KLISS, BNRist, School of Software, Tsinghua University, China}
}
\email{jiangyu198964@126.com}

\author{Yue Zhao}
\affiliation{%
  \institution{KLISS, BNRist, School of Software, Tsinghua University, China}
}
\email{zhao-y17@mails.tsinghua.edu.cn}

\author{Quan Chen}
\affiliation{%
  \institution{Department of Computer Science and Engineering, Shanghai Jiao Tong University, China}
}
\email{chen-quan@cs.sjtu.edu.cn}

\author{Jiaguang Sun}
\affiliation{%
  \institution{KLISS, BNRist, School of Software, Tsinghua University, China}
}
\email{sunjg@tsinghua.edu.cn}

\begin{abstract}
Deep learning (DL) systems are increasingly applied to safety-critical domains such as autonomous driving cars. It is of significant importance to ensure the reliability and robustness of DL systems. Existing testing methodologies always fail to include rare inputs in the testing dataset and exhibit low neuron coverage. 

In this paper, we propose DLFuzz, the first differential fuzzing testing framework to guide DL systems exposing incorrect behaviors. DLFuzz keeps minutely mutating the input to maximize the neuron coverage and the prediction difference between the original input and the mutated input, without manual labeling effort or cross-referencing oracles from other DL systems with the same functionality. We present empirical evaluations on two well-known datasets to demonstrate its efficiency. Compared with DeepXplore, the state-of-the-art DL whitebox testing framework, DLFuzz does not require extra efforts to find similar functional DL systems for cross-referencing check, but could generate $338.59\%$ more adversarial inputs with $89.82\%$ smaller perturbations, averagely obtain $2.86\%$ higher neuron coverage, and save $20.11\%$ time consumption.
\end{abstract}

\ccsdesc[500]{Software and its engineering~Software testing and debugging}

\keywords{Fuzzing Testing, Deep Learning, Neuron Coverage}

\maketitle

\section{Introduction}\label{sec:introduction} 
In the past few years, deep learning~(DL) systems have demonstrated its competitiveness on a wide range of applications, such as image classification~\cite{krizhevsky2012imagenet,he2016deep}, natural language processing~\cite{sutskever2014sequence} and even reconstruction of brain circuits~\cite{helmstaedter2013connectomic}. These encouraging accomplishments inspired wide deployments of DL systems in safety-critical domains, such as autonomous driving~\cite{bojarski2016end}, drones and robotics~\cite{mnih2015human} and malware detection~\cite{yuan2014droid}, and it is in great demand to test and improve the robustness of DL systems.

\begin{figure}[!htbp]
\centering
\setlength{\belowcaptionskip}{-10pt}
\includegraphics[width=0.46\textwidth]{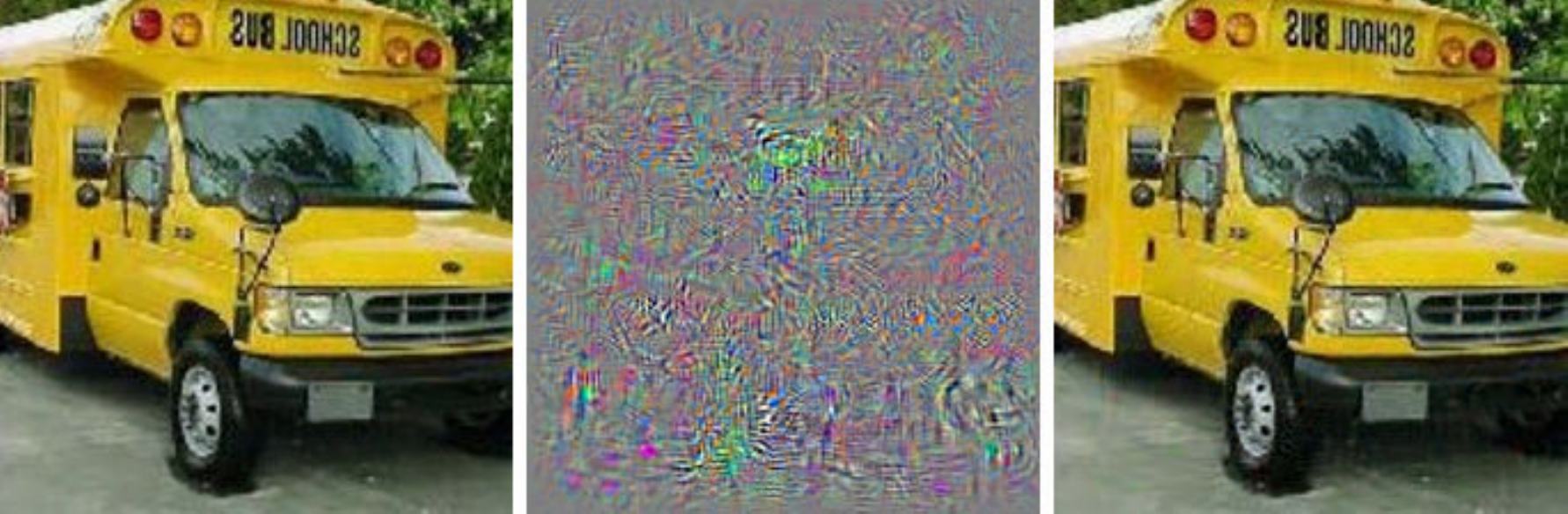}
\caption{An adversarial input for DL model AlexNet\cite{krizhevsky2012imagenet}. (Left) is correctly predicted as \textit{"school bus"}, (center) perturbation applied to the left to obtain the right, (right) adversarial input predicted as \textit{"ostrich"}.}
\label{fig:introduction_exp}
\end{figure}

For DL testing, the classical approach is to gather sufficient manually labeled data to assess the accuracy of DL systems. However, the input space of testing is so huge that it is extremely hard to collect all the possible inputs to trigger every feasible logic of a DL system. It is demonstrated that state-of-the-art DL systems can be fooled by adding small perturbations to the test inputs~\cite{szegedy2013intriguing}. Although DL systems exhibit impressive performance on image classification tasks, the classifiers can also be easily led to incorrect classifications by applying imperceptible perturbations~\cite{moosavi2016deepfool}, as shown in Figure \ref{fig:introduction_exp}. Therefore, DL testing is quite challenging but essential to ensure the correctness of those safety-critical practices.

Several approaches have been proposed to improve the testing efficiency of DL systems. Some of them leverage solvers like Z3 to generate adversarial inputs under the formalized constraints of the DL models~\cite{huang2017safety, gu2014towards}. These techniques are accurate, but work in a heavy whitebox manner and are resource-consuming for constraint solving. Some blackbox methods exploit heuristic algorithms to mutate the inputs until the adversarial inputs acquired~\cite{wicker2018feature}. These methods are time-consuming and rely heavily on the manually supplied ground truth labels. Other approaches of adversarial deep learning focus on fooling the DL systems by applying imperceptible perturbations to the inputs mostly in a gradient-based manner~\cite{szegedy2013intriguing, moosavi2016deepfool}. They work efficiently but are shown to have low neuron coverage~\cite{pei2017deepxplore}. Recently, DeepXplore~\cite{pei2017deepxplore} was presented as the state-of-the-art whitebox testing framework for DL systems and first introduced the concept of neuron coverage as a testing metric. Meanwhile, it requires multiple DL systems with similar functionality as cross-referencing oracles to avoid manual checking. Nevertheless, cross-referencing suffers from the scalability and difficulty of finding similar DL systems. 

In this paper, we propose DLFuzz\footnote{https://github.com/turned2670/DLFuzz}, the first differential fuzzing testing framework aiming to maximize the neuron coverage and generate more adversarial inputs for a given DL system, without cross-referencing other similar DL systems or manual labeling. First, DLFuzz iteratively selects neurons worthy to activate for covering more logic, and mutates the test inputs by applying minute perturbations to guide the DL systems exposing incorrect behaviors. 
During the mutation process, DLFuzz keeps the mutated inputs which contribute to a certain increase of the neuron coverage for the subsequent fuzzing, and the minute perturbation is restricted to be invisible and ensures that the prediction results of the original input and the mutated inputs should be the same.
In this way, DLFuzz is able to obtain rare inputs, 
and can automatically identify the erroneous behaviors with differential testing that an error is triggered when the prediction result of the mutated input is not the same with the original input. 

To evaluate the efficiency of DLFuzz, we conducted empirical studies on six DL systems trained on two popular datasets, MNIST~\cite{lecun1998mnist} and ImageNet~\cite{deng2009imagenet}. The DL systems and the datasets are exactly the same as those used by DeepXplore. Compared with DeepXplore, DLFuzz does not need extra efforts to collect similar functional DL systems for cross-referencing label check, but could generate $135\%-584.62\%$ more adversarial inputs with $79.56\%-96.77\%$ smaller perturbations, and obtain $1.10\%-5.59\%$ higher neuron coverage. For the time efficiency, it saves $20.11\%$ time consumption in average with one exceptional case on ImageNet costing $59.42\%$ more than DeepXplore.

\section{Motivation}\label{sec:motivation}
There exists a large gap between DL testing and traditional software testing, owing to the totally distinct internal structures of deep neural networks~(DNN) and software programs. 
Researchers have devoted many efforts to applying software testing to DL testing in both whitebox and blackbox manner~\cite{huang2017safety, gu2014towards, wicker2018feature}. 
We aim to break the resource consumption limitation of blackbox testing techniques\cite{wicker2018feature} and the cross-referencing obstacles of whitebox testing techniques\cite{pei2017deepxplore} with DLFuzz, the first  differential fuzzing testing framework leveraging imperceptible perturbations to ensure the invisible difference between the inputs and the mutated inputs.

Fuzzing testing~\cite{zalewski2007american,wang2018safl,liang2018fuzz} has been recognized as one of the most effective methodologies for vulnerability detection in software testing, demonstrated by the huge amount of vulnerabilities caught. The core idea is to generate random inputs to execute as many program paths as possible so as to lead the program to expose violations and crashes. It can be seen that fuzzing testing and DL testing share similar goals of achieving higher coverage as well as getting more exceptional behaviors. In general, we combined the knowledge in key stages of fuzzing into DL testing as below:
\begin{enumerate}
\item \textbf{Optimization Goal.} The goal of reaching higher neuron coverage and exposing more exceptional behaviors can be treated as a joint optimization problem. This optimization problem can be implemented in the gradient-based manner.
\item \textbf{Seed Maintenance.} While fuzzing, the mutated inputs which contribute to a certain increase of the neuron coverage are kept in the seed list, based on the potential to improve neuron coverage continuously in the subsequent fuzzing.
\item \textbf{Diversity in Mutation Strategies.} We designed many neuron selection strategies to select neurons that are possible to cover more logic and trigger more incorrect outputs. Furthermore, multiple mutation ways for test inputs have been already practiced, and are easy to be integrated. 
\end{enumerate}

\section{DLFuzz Approach}\label{sec:approach}
\subsection{Architecture}
\begin{figure}[!htbp]
\setlength{\belowcaptionskip}{-2pt}
\includegraphics[width=0.48\textwidth]{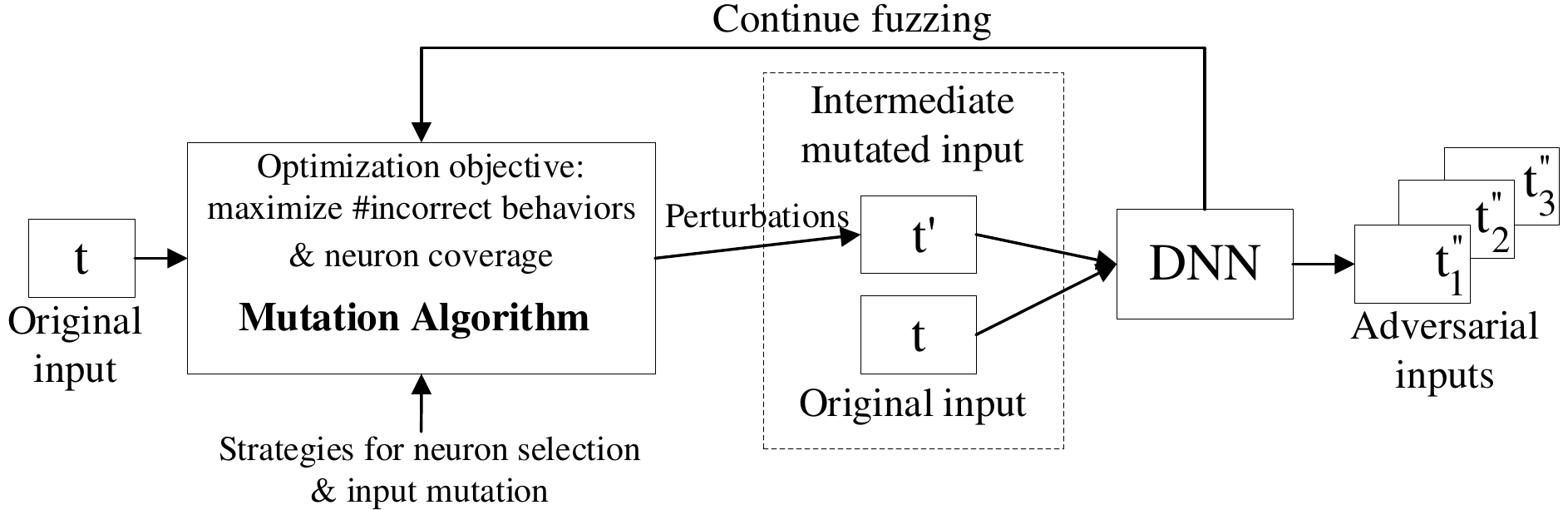}
\caption{Architecture of DLFuzz}
\label{fig:architecture}
\end{figure}

The overall architecture of DLFuzz is depicted in Figure \ref{fig:architecture}. In this paper, we implement DLFuzz to work on image classification, a popular task in DL domains to demonstrate its feasibility and effectiveness. The adaptions in other tasks such as speech recognition are straightforward and also follow the same workflow in Figure \ref{fig:architecture}.

To specify, the given test input $t$ is an image to be classified, the DNN is a particular convolutional neural network~(CNN) under test, such as VGG-16~\cite{simonyan2014very}. The mutation algorithm applies tiny perturbation to $t$ and gets $t^\prime$, which is visibly indistinguishable from $t$. If the mutated input $t^\prime$ and the original input $t$ are both fed to the CNN but classified to be of different class labels, we treat this as an incorrect behavior and $t^\prime$ to be one of the adversarial inputs. The inconsistent classification results before and after mutation indicate that at least one of them is wrong so that manually labeling effort is not required here. In contrast, if the two are predicted of the same class label by the CNN, $t^\prime$ will continue to be mutated by the mutation algorithm to test the CNN's robustness. 

\subsection{Algorithm}
The mutation algorithm is the main component of DLFuzz. It is completed by solving a joint optimization problem of both maximizing the neuron coverage and the number of incorrect behaviors. Based on the demonstration that covering more neurons could potentially trigger more logic and more erroneous behaviors~\cite{pei2017deepxplore}, DLFuzz also leverages the same definition and computing way of neuron coverage as DeepXplore~\cite{pei2017deepxplore}. Neurons with output values larger than the set threshold are regarded as activated~(covered).

The core process of the mutation algorithm is in Algorithm \ref{alg:algorithm}. The algorithm contains three key components to discuss in detail.

\setlength{\textfloatsep}{0.1cm}
\setlength{\floatsep}{0.1cm}
\begin{algorithm}[!htbp]
\caption{Mutation Algorithm}\label{alg:algorithm}
\begin{algorithmic}[1]
\REQUIRE input\_list $\leftarrow$ unlabeled inputs for testing\\
	 	 \quad\ \ dnn $\leftarrow$ DNN under test\\
	     \quad\ \ k $\leftarrow$ top k labels different from the original label\\
	     \quad\ \ m $\leftarrow$ number of neurons to cover\\
	     \quad\ \ strategies $\leftarrow$ strategies for neuron selection\\
	     \quad\ \ $\lambda \leftarrow$ hyperparameter for balancing two goals\\
	     \quad\ \ cov\_tracker $\leftarrow$ tracks information of neurons\\
	     \quad\ \ iter\_times $\leftarrow$ iteration times for each seed
\ENSURE set of adversarial inputs, neuron coverage
\STATE adversarial\_set = []
\FOR{x in input\_list}
\STATE seed\_list = [x] \COMMENT{\textit{\small{seeds for each input}}}
\WHILE{len(seed\_list) > 0} 
\STATE $x_s$ = seed\_list[0] \COMMENT{\textit{\small{grab the head element}}}
\STATE seed\_list.remove($x_s$)
\STATE c, c\_topk = dnn.predict($x_s$)
\STATE neurons = selection(dnn, cov\_tracker, strategies, m)
\STATE obj = sum(c\_topk) - c + $\lambda$ $\cdot$ sum(neurons)
\STATE grads = $\partial obj / \partial x_s$ \COMMENT \textit{\small{gradient obtained}}
\FOR{iter=0 to iter\_times}
\STATE /*\textit{\small{gradient processed to get the perturbation for mutation}}*/
\STATE perturbation = processing(grads) 
\STATE $x^\prime$ = $x_s$ + perturbation \COMMENT \textit{\small{mutated input obtained}}
\STATE $c^\prime$ = dnn.predict($x^\prime$) \COMMENT \textit{\small{label after mutation}}
\STATE update cov\_tracker \COMMENT \textit{\small{update coverage information}}
\STATE $l_2\_distance$ = distance($x^\prime$, x) \COMMENT \textit{\small{measure the perturbation}}
\IF{coverage improved by $x^\prime$ is desired and  $l_2\_distance$ is small}
\STATE seed\_list.append($x^\prime$)
\ENDIF
\IF{$c^\prime$ != c}
\STATE adversarial\_set.append($x^\prime$)
\STATE \textbf{break}
\ENDIF
\ENDFOR
\ENDWHILE
\ENDFOR
\end{algorithmic}
\end{algorithm}

\textbf{Optimization Problem.} As discussed in Section \ref{sec:introduction}, the gradient-based adversarial deep learning outperforms the other approaches in several aspects, especially in time efficiency. It founds perturbations by optimizing the input to maximize the prediction error~\cite{szegedy2013intriguing}, which is opposite to optimizing the weights to minimize the prediction error while training the DNN. It is easy to implement by customizing the loss function as our objective and maximizing the loss by gradient ascent.

The loss function of DLFuzz is defined as the following equation~(Algorithm \ref{alg:algorithm} line 9), which is also the optimization objective:
\begin{equation}
  obj = \sum_{i=0}^{k} c_i - c + \lambda \cdot \sum_{i=0}^{m} n_i
\end{equation}
where the objective consists of two parts. In the first part $\sum_{i=0}^{k} c_i - c$, $c$ is the original class label of the input, $c_i(i=0,...,k)$ is one of top k class labels with confidence just lower than $c$~(Algorithm \ref{alg:algorithm} line 7). Maximizing the first part guides the input to cross the decision boundary of the original class and lie in the decision space of top k other classes. Such modified inputs are more likely to be classified incorrectly~\cite{moosavi2016deepfool}. In the second part $\sum_{i=0}^{m} n_i$,  $n_i$ is a target neuron intended to activate. These neurons are selected considering many strategies to improve neuron coverage~(Algorithm \ref{alg:algorithm} line 8). The hyperparameter $\lambda$ is used for balancing the two objectives. 

\textbf{Fuzzing Process.} The fuzzing process reveals the overall workflow of Algorithm \ref{alg:algorithm}. When given a test input $x$, DLFuzz maintains a seed list for keeping the intermediate mutated inputs which contribute to neuron coverage. At first, the seed list only has one input which is exactly $x$. Next, DLFuzz traverses each seed $x_s$ and obtains the elements making up its optimization objective. Then, DLFuzz computes the gradient direction for later mutation. In the mutation process, DLFuzz iteratively applies the processed gradient as the perturbation to $x_s$ and obtains the intermediate input $x^\prime$. After each mutation, the intermediate class label $c^\prime$, coverage information, $l_2$ distance of $x$ and $x^\prime$ are acquired. If the neuron coverage improved by $x^\prime$ and $l_2$ distance are desired, $x^\prime$ will be added into the seed list. Finally, if $c^\prime$ is already different from $c$, mutation process for seed $x_s$ terminates and $x^\prime$ will be included in the set of adversarial inputs. Therefore, DLFuzz is able to generate multiple adversarial inputs for a certain original input and explore a new way to further improve neuron coverage. 

For the iterative mutation process, first, various processing methods are available to generate perturbations when the gradients obtained, including just keeping the sign~\cite{szegedy2013intriguing}, imitating the realistic situations~\cite{tian2018deeptest,pei2017deepxplore}, etc. These mutation strategies for the input are easy to be extended to DLFuzz. Second, DLFuzz adopts $l_2$ distance to measure the perturbation with the same computation as ~\cite{moosavi2016deepfool}, so as to ensure the distance between $x$ and $x^\prime$ is imperceptible. As for the conditions of seed keeping in line 18, DLFuzz limits our desired distance to a relatively small range~(less than 0.02) to ensure the imperceptibility. As the neuron coverage improvement of one input declines with time, the corresponding threshold for keeping the seed also decreases with running time. In addition, the inputs and hyperparameters configuring DLFuzz have certain effects to the performance and need some exploration efforts. Furthermore, we can increase the thresholds of seed keeping to reserve more mutated inputs with greater distance.

\textbf{Strategies for Neuron Selection.} To maximize neuron coverage, we propose four heuristic strategies for selecting neurons more likely to improve coverage as listed below. For each seed $x_s$, $m$ neurons will be selected utilizing one or multiple strategies, which can be customized in $strategies$ of the algorithm inputs.
\begin{enumerate}[topsep=3pt]
\item Strategy 1. Select neurons covered frequently during past testing. Inspired by practical experience in traditional software testing that code fragments often or rarely executed are more possible to introduce defects. Neurons covered often or rarely perhaps can result in unusual logic and activate more neurons.
\item Strategy 2. Select neurons covered rarely during past testing due to the considerations stated above. 
\item Strategy 3. Select neurons with top weights. It is presented based on our assumption that neurons with top weights maybe have larger influence on other neurons. 
\item Strategy 4. Select neurons near the activation threshold. It is easier to accelerate if activating/deactivating neurons with output value slightly lower/larger than the threshold.
\end{enumerate}

\section{Experiment}\label{sec:experiment}
\subsection{Experiment Setup}
We implemented DLFuzz based on 
the widespread frameworks of DL systems, Tensorflow 1.2.1 and Keras 2.1.3. Tensorflow and Keras provide the efficient interfaces for gradient computations and support the process of recording the intermediate output of all neurons after each prediction of the DNN. We developed and evaluated DLFuzz on a computer with 4 cores~(Intel i7-7700HQ @3.6GHz), 16GB of memory, a NVIDIA GTX 1070 GPU and Ubuntu 16.04.4 as the host OS. 

We selected two datasets~(MNIST and ImageNet) and the corresponding CNNs used by DeepXplore for image classification tasks to evaluate DLFuzz. MNIST~\cite{lecun1998mnist} is a large database of handwritten digits consisting of 60000 training images and 10000 testing images. ImageNet~\cite{deng2009imagenet} is a large visual database containing over 14 million images for object recognition. The same as DeepXplore, DLFuzz tested three pre-trained models for each dataset, that is, LeNet-1~\cite{lecun1998gradient}, LeNet-4~\cite{lecun1998gradient}, LeNet-5~\cite{lecun1998gradient} for MNIST and VGG-16~\cite{simonyan2014very}, VGG-19~\cite{simonyan2014very}, ResNet50~\cite{he2016deep} for ImageNet. Considering the fairness, we also randomly choose 20 images from the dataset for each CNN as test inputs in the same way with DeepXplore. As for the hyperparameters configured in the input, we tried lots of combinations of possible settings. If not specified, the default settings of those hyperparameters $k$, $m$, $strategies$, $iter\_times$ are $4, 10, "strategy\ 1", 3$ respectively for their good performance on many DNNs. 

\subsection{Result}
Table \ref{table:main table} presents the effectiveness of DLFuzz compared with DeepXplore. DLFuzz exhibits its advantages in improving neuron coverage, generating more adversarial inputs within the same time limit and restricting imperceptible  perturbations. First, as presented in the third column of the table, for the tested six CNNs, DLFuzz achieves $1.10\%$ to $5.59\%$ higher neuron coverage than DeepXplore in different settings in average. For the best setting, DLFuzz is able to acquire $13.42\%$ higher neuron coverage. 

\begin{table}[htbp]
\caption{Effectiveness of DLFuzz compared with DeepXplore.} 
\label{table:main table}
\begin{threeparttable}
\resizebox{8.5cm}{1.6cm}{
\begin{tabular}{p{1.0cm}p{1.8cm}p{0.6cm}p{1.5cm}p{0.6cm}p{1.7cm}}
\toprule
DataSet  & Model\quad(\#Neurons) & NC Imp.\tnote{1} & $l_2$ Distance & \#Adv.\tnote{2} & Adv. Time\tnote{3} \\ \midrule
MNIST    & LeNet-1(52)      & 2.45\%                              & 8.2637/0.2708                  & 20/53                           & 0.7078/0.5623                                               \\
         & LeNet-4(148)     & 5.59\%                              & 8.2637/0.2812                  & 20/47                           & 0.7078/0.6344                                                \\
         & LeNet-5(268)     & 2.23\%                              & 8.2637/0.2670                  & 20/54                           & 0.7078/0.5870                                                \\
ImageNet & VGG16(14888)     & 3.52\%                              & 0.0817/0.0167                  & 13/89                           & 10.473/3.4537                                                \\
         & VGG19(16168)     & 2.28\%                              & 0.0817/0.0154                  & 13/81                           & 10.473/3.6606                                                \\
         & ResNet50(94056)  & 1.10\%                              & 0.0817/0.0097                  & 13/72                           & 10.473/16.6958                                               \\ \bottomrule
\end{tabular}}
\begin{tablenotes}
\footnotesize
\item Comparisons represented by content in format a/b, where a denotes the result\\of DeepXplore and b denotes the result of DLFuzz.
\item[1] Average neuron coverage improvement.
\item[2] Number of adversarial inputs generated.
\item[3] Average time of generating per adversarial input.
\end{tablenotes}
\end{threeparttable}
\end{table}

\begin{figure}[!htbp]
\centering
\begin{minipage}[!htbp]{0.11\textwidth}
     \centering
     \includegraphics[width=1.0\textwidth]{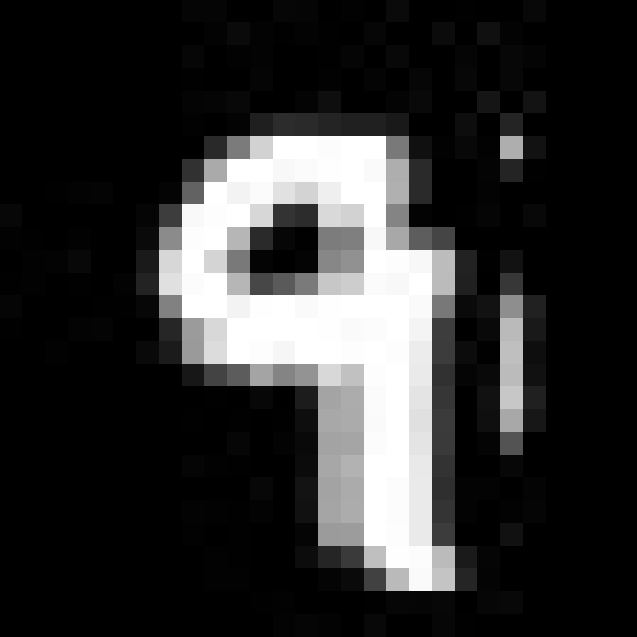}
     \small{Original: 9}
\end{minipage}
\begin{minipage}[!htbp]{0.11\textwidth}
     \centering
     \includegraphics[width=1.0\textwidth]{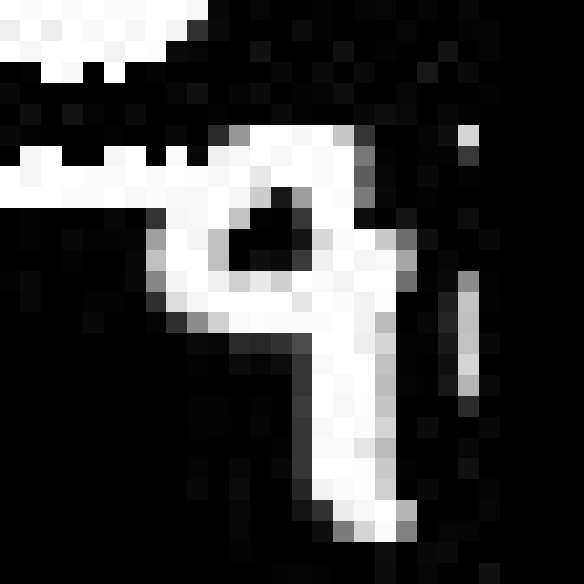}     
     \small{1-DeepXplore:4}
     \end{minipage}
\begin{minipage}[!htbp]{0.11\textwidth}
     \centering
     \includegraphics[width=1.0\textwidth]{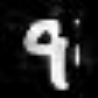}     
     \small{1-DLFuzz: 4}
\end{minipage}
\begin{minipage}[!htbp]{0.11\textwidth}
     \centering
     \includegraphics[width=1.0\textwidth]{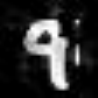}     
     \small{2-DLFuzz: 4}
\end{minipage}
\begin{minipage}[!htbp]{0.11\textwidth}
     \centering
     \includegraphics[width=1.0\textwidth]{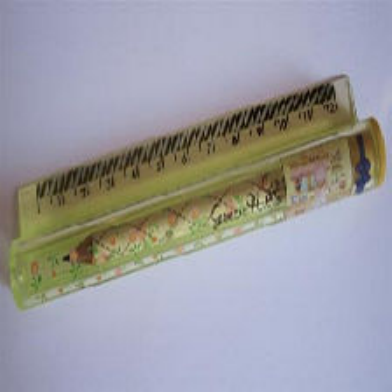}
     \small{Original:\\rule}
\end{minipage}
\begin{minipage}[!htbp]{0.11\textwidth}
     \centering
     \includegraphics[width=1.0\textwidth]{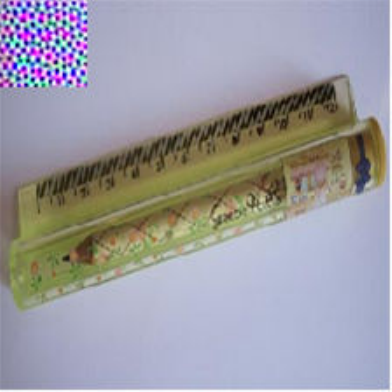}     
     \small{1-DeepXplore:\\harmonica}
\end{minipage}
\begin{minipage}[!htbp]{0.11\textwidth}
     \centering
     \includegraphics[width=1.0\textwidth]{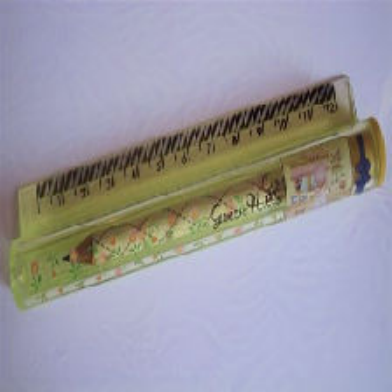}     
     \small{1-DLFuzz:\\paper}
\end{minipage}
\begin{minipage}[!htbp]{0.11\textwidth}
     \centering
     \includegraphics[width=1.0\textwidth]{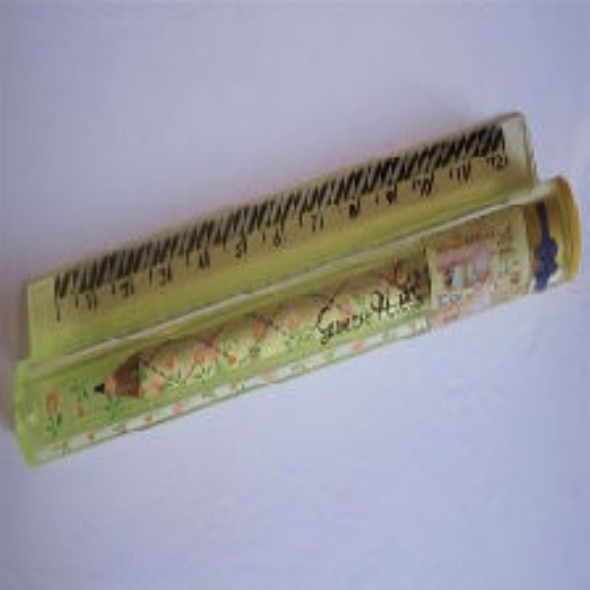}     
     \small{2-DLFuzz:\\pencil box}
\end{minipage}
\caption{Cases of adversarial inputs annotated with the framework and the predicted label. The above row for MNIST and the below for ImageNet.} 
\label{fig:example}
\end{figure}

Next, adversarial inputs generated by DLFuzz have much smaller perturbations. As the cases in Figure \ref{fig:example}, the perturbations generated by DeepXplore are visible while those generated by DLFuzz are invisible and imperceptible. In this way, DLFuzz provides stronger guarantee for the consistence of the image's identity before and after mutation. As shown in the fourth and fifth column of table 1, DLFuzz averagely generated $338.59\%$ more adversarial inputs with $89.82\%$ smaller perturbations. Moreover, DLFuzz spent $20.11\%$ shorter time on generating each adversarial input on these DL systems. An exceptional case is that DLFuzz cost more time on generating  adversarial inputs than DeepXplore for ResNet50, which is owing to more time needed for neuron selection when testing a DL system consisting of a huge amount of neurons~(94056). 

\begin{figure}[htbp]
\centering
\begin{minipage}[!htbp]{0.23\textwidth}
     \centering
     \includegraphics[width=1.0\textwidth]{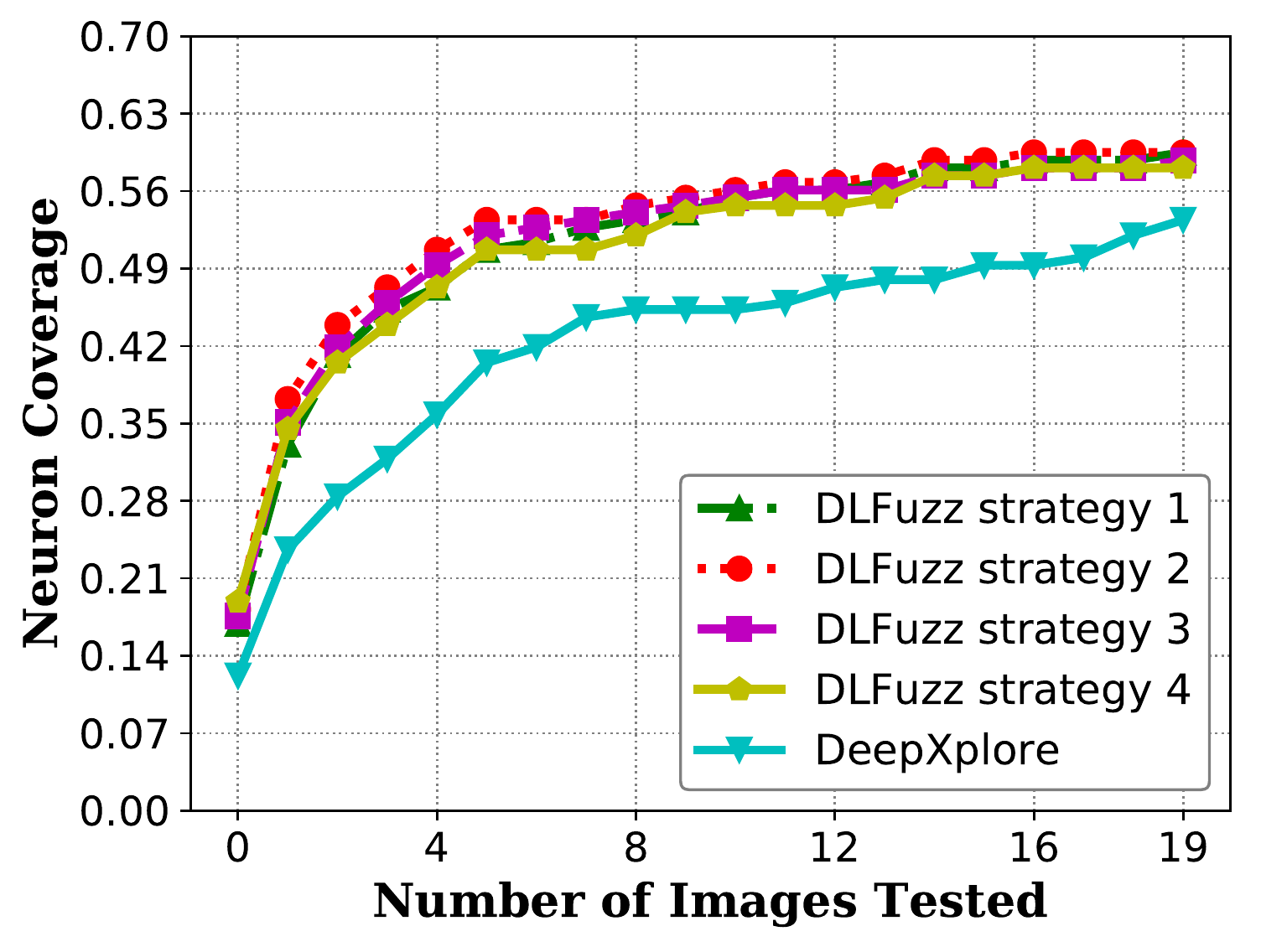}
     \small{LeNet-4}
\end{minipage}
\begin{minipage}[!htbp]{0.23\textwidth}
     \centering
     \includegraphics[width=1.0\textwidth]{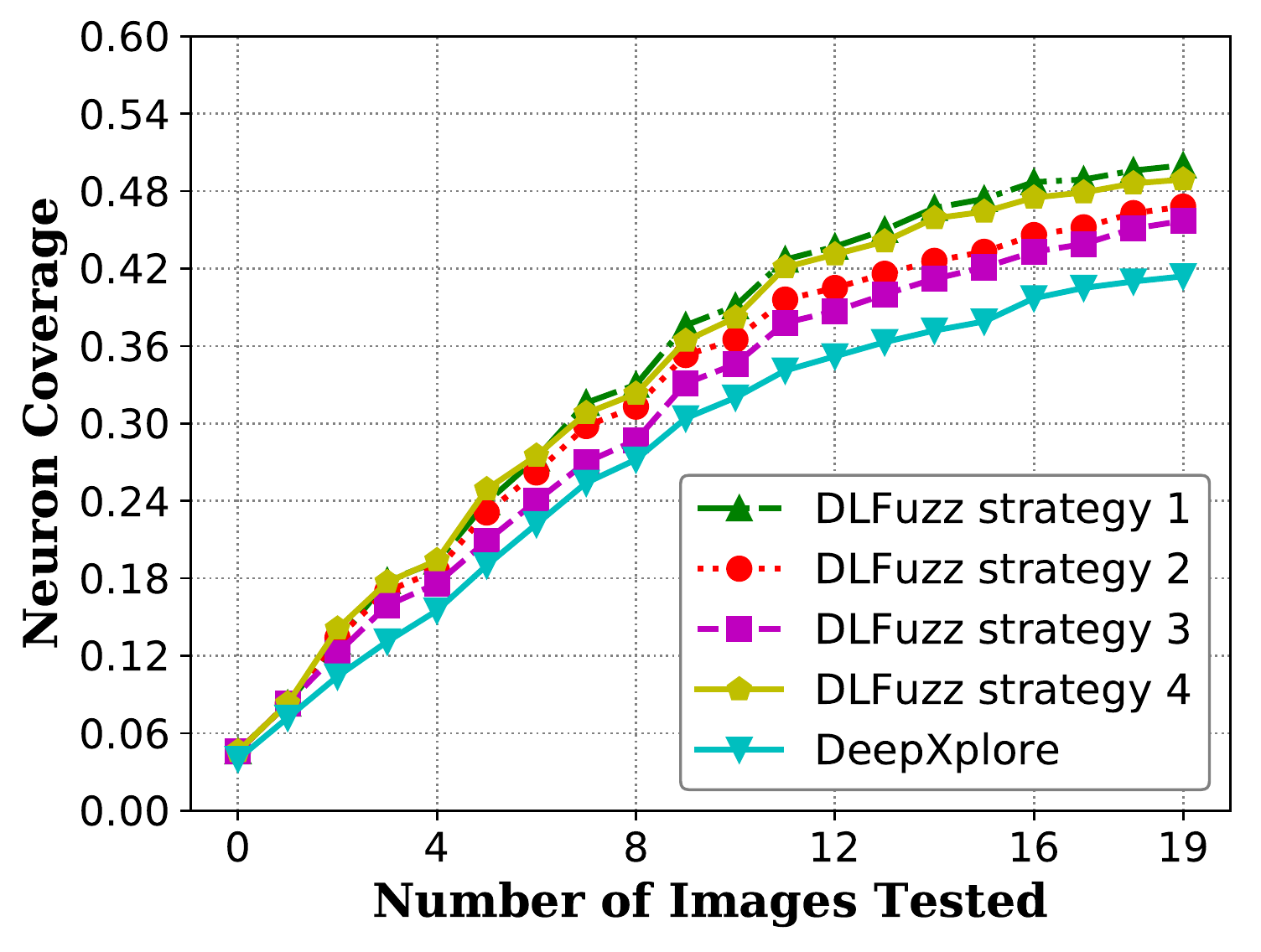}     
     \small{VGG16}
\end{minipage}
\caption{Neuron coverage with number of images tested when different strategies applied in DLFuzz. }
\label{fig:strategies}
\end{figure}

We also tried all the proposed neuron selection strategies on two CNNs and depicted the results in Figure \ref{fig:strategies}. All strategies are shown to contribute more to neuron coverage improvement than DeepXplore while have similar performance among themselves. It seems that "strategy 1" performs slightly better.
In addition, to prove the practical use of DLFuzz, we incorporated 114 adversarial images into the training set of three CNNs on MNIST and retrained them to see if their accuracy is able to increase. Finally, we improve their accuracy by up to $1.8\%$ within 5 epochs. More improvement is expected if more adversarial inputs included in the retraining process.

\subsection{Discussion}
\textbf{Applicability of Fuzzing to DL Testing.} The effectiveness of DLFuzz demonstrates that applying the knowledge of fuzzing to DL testing is feasible and can greatly improve the performance of existing DL testing techniques such as DeepXplore. The gradient-based solution of the optimization problem guarantees the easy deployment and high efficiency of the framework. The mechanism of seed maintenance provides diverse directions and larger space for improving neuron coverage. As the cases shown in Figure \ref{fig:example}, it is also capable to obtain incremental adversarial inputs for one input. Various strategies combined for neuron selection proved to be good at finding neurons beneficial for increasing neuron coverage.

\textbf{Without Manual Effort.} For confirmation, we checked all the 366 adversarial inputs generated by DLFuzz, though DLFuzz maintains quite small $l_2$ distance by the restricted threshold. We haven't found any adversarial inputs that have already changed their identities after mutation. The adversarial inputs are nearly the same as the original input, and the perturbations are imperceptible. 

\textbf{Future Work.} Encouraged by the impressive effects of DLFuzz on image classification tasks, we will work on the deployments of DLFuzz on other popular tasks in DL domains, such as speech recognition. The specific constraints for input mutation of the corresponding task will be added into the common workflow. Also, some domain knowledge can be leveraged to provide more efficient mutation operations and increase the the efficiency of DLFuzz.

\section{Conclusion}\label{sec:conclusions}
\balance
We design and implement DLFuzz as an effective fuzzing testing framework of DL systems. DLFuzz first combines the basic ideas of fuzzing testing into DL testing and demonstrates its efficiency. Compared with DeepXplore, DLFuzz averagely obtained $2.86\%$ higher neuron coverage and generated $338.59\%$ more adversarial examples with $89.82\%$ smaller perturbations given the same amount of inputs. DLFuzz also overcomes the trouble of relying on multiple DL systems of the similar functionality in DeepXplore. Additionally, DLFuzz exhibits its practical use by incorporating these adversarial inputs to retrain the DL systems and to steadily improve their accuracy. 
The premiere results present its potential usage to expose the incorrect behaviors of DL systems at an early stage, and ensure the reliability and robustness.

\bibliographystyle{ACM-Reference-Format}
\bibliography{sample-bibliography}


\begin{thebibliography}{21}


\ifx \showCODEN    \undefined \def \showCODEN     #1{\unskip}     \fi
\ifx \showDOI      \undefined \def \showDOI       #1{#1}\fi
\ifx \showISBNx    \undefined \def \showISBNx     #1{\unskip}     \fi
\ifx \showISBNxiii \undefined \def \showISBNxiii  #1{\unskip}     \fi
\ifx \showISSN     \undefined \def \showISSN      #1{\unskip}     \fi
\ifx \showLCCN     \undefined \def \showLCCN      #1{\unskip}     \fi
\ifx \shownote     \undefined \def \shownote      #1{#1}          \fi
\ifx \showarticletitle \undefined \def \showarticletitle #1{#1}   \fi
\ifx \showURL      \undefined \def \showURL       {\relax}        \fi
\providecommand\bibfield[2]{#2}
\providecommand\bibinfo[2]{#2}
\providecommand\natexlab[1]{#1}
\providecommand\showeprint[2][]{arXiv:#2}

\bibitem[\protect\citeauthoryear{Bojarski, Del~Testa, Dworakowski, Firner,
  Flepp, Goyal, Jackel, Monfort, Muller, Zhang, et~al\mbox{.}}{Bojarski
  et~al\mbox{.}}{2016}]%
        {bojarski2016end}
\bibfield{author}{\bibinfo{person}{Mariusz Bojarski}, \bibinfo{person}{Davide
  Del~Testa}, \bibinfo{person}{Daniel Dworakowski}, \bibinfo{person}{Bernhard
  Firner}, \bibinfo{person}{Beat Flepp}, \bibinfo{person}{Prasoon Goyal},
  \bibinfo{person}{Lawrence~D Jackel}, \bibinfo{person}{Mathew Monfort},
  \bibinfo{person}{Urs Muller}, \bibinfo{person}{Jiakai Zhang},
  {et~al\mbox{.}}} \bibinfo{year}{2016}\natexlab{}.
\newblock \showarticletitle{End to end learning for self-driving cars}.
\newblock \bibinfo{journal}{\emph{arXiv preprint arXiv:1604.07316}}
  (\bibinfo{year}{2016}).
\newblock


\bibitem[\protect\citeauthoryear{Deng, Dong, Socher, Li, Li, and Fei-Fei}{Deng
  et~al\mbox{.}}{2009}]%
        {deng2009imagenet}
\bibfield{author}{\bibinfo{person}{Jia Deng}, \bibinfo{person}{Wei Dong},
  \bibinfo{person}{Richard Socher}, \bibinfo{person}{Li-Jia Li},
  \bibinfo{person}{Kai Li}, {and} \bibinfo{person}{Li Fei-Fei}.}
  \bibinfo{year}{2009}\natexlab{}.
\newblock \showarticletitle{Imagenet: A large-scale hierarchical image
  database}. In \bibinfo{booktitle}{\emph{Computer Vision and Pattern
  Recognition, 2009. CVPR 2009. IEEE Conference on}}. IEEE,
  \bibinfo{pages}{248--255}.
\newblock


\bibitem[\protect\citeauthoryear{Gu and Rigazio}{Gu and Rigazio}{2014}]%
        {gu2014towards}
\bibfield{author}{\bibinfo{person}{Shixiang Gu} {and} \bibinfo{person}{Luca
  Rigazio}.} \bibinfo{year}{2014}\natexlab{}.
\newblock \showarticletitle{Towards deep neural network architectures robust to
  adversarial examples}.
\newblock \bibinfo{journal}{\emph{arXiv preprint arXiv:1412.5068}}
  (\bibinfo{year}{2014}).
\newblock


\bibitem[\protect\citeauthoryear{He, Zhang, Ren, and Sun}{He
  et~al\mbox{.}}{2016}]%
        {he2016deep}
\bibfield{author}{\bibinfo{person}{Kaiming He}, \bibinfo{person}{Xiangyu
  Zhang}, \bibinfo{person}{Shaoqing Ren}, {and} \bibinfo{person}{Jian Sun}.}
  \bibinfo{year}{2016}\natexlab{}.
\newblock \showarticletitle{Deep residual learning for image recognition}. In
  \bibinfo{booktitle}{\emph{Proceedings of the IEEE conference on computer
  vision and pattern recognition}}. \bibinfo{pages}{770--778}.
\newblock


\bibitem[\protect\citeauthoryear{Helmstaedter, Briggman, Turaga, Jain, Seung,
  and Denk}{Helmstaedter et~al\mbox{.}}{2013}]%
        {helmstaedter2013connectomic}
\bibfield{author}{\bibinfo{person}{Moritz Helmstaedter},
  \bibinfo{person}{Kevin~L Briggman}, \bibinfo{person}{Srinivas~C Turaga},
  \bibinfo{person}{Viren Jain}, \bibinfo{person}{H~Sebastian Seung}, {and}
  \bibinfo{person}{Winfried Denk}.} \bibinfo{year}{2013}\natexlab{}.
\newblock \showarticletitle{Connectomic reconstruction of the inner plexiform
  layer in the mouse retina}.
\newblock \bibinfo{journal}{\emph{Nature}} \bibinfo{volume}{500},
  \bibinfo{number}{7461} (\bibinfo{year}{2013}), \bibinfo{pages}{168}.
\newblock


\bibitem[\protect\citeauthoryear{Huang, Kwiatkowska, Wang, and Wu}{Huang
  et~al\mbox{.}}{2017}]%
        {huang2017safety}
\bibfield{author}{\bibinfo{person}{Xiaowei Huang}, \bibinfo{person}{Marta
  Kwiatkowska}, \bibinfo{person}{Sen Wang}, {and} \bibinfo{person}{Min Wu}.}
  \bibinfo{year}{2017}\natexlab{}.
\newblock \showarticletitle{Safety verification of deep neural networks}. In
  \bibinfo{booktitle}{\emph{International Conference on Computer Aided
  Verification}}. Springer, \bibinfo{pages}{3--29}.
\newblock


\bibitem[\protect\citeauthoryear{Krizhevsky, Sutskever, and Hinton}{Krizhevsky
  et~al\mbox{.}}{2012}]%
        {krizhevsky2012imagenet}
\bibfield{author}{\bibinfo{person}{Alex Krizhevsky}, \bibinfo{person}{Ilya
  Sutskever}, {and} \bibinfo{person}{Geoffrey~E Hinton}.}
  \bibinfo{year}{2012}\natexlab{}.
\newblock \showarticletitle{Imagenet classification with deep convolutional
  neural networks}. In \bibinfo{booktitle}{\emph{Advances in neural information
  processing systems}}. \bibinfo{pages}{1097--1105}.
\newblock


\bibitem[\protect\citeauthoryear{LeCun}{LeCun}{1998}]%
        {lecun1998mnist}
\bibfield{author}{\bibinfo{person}{Yann LeCun}.}
  \bibinfo{year}{1998}\natexlab{}.
\newblock \showarticletitle{The MNIST database of handwritten digits}.
\newblock \bibinfo{journal}{\emph{http://yann. lecun. com/exdb/mnist/}}
  (\bibinfo{year}{1998}).
\newblock


\bibitem[\protect\citeauthoryear{LeCun, Bottou, Bengio, and Haffner}{LeCun
  et~al\mbox{.}}{1998}]%
        {lecun1998gradient}
\bibfield{author}{\bibinfo{person}{Yann LeCun}, \bibinfo{person}{L{\'e}on
  Bottou}, \bibinfo{person}{Yoshua Bengio}, {and} \bibinfo{person}{Patrick
  Haffner}.} \bibinfo{year}{1998}\natexlab{}.
\newblock \showarticletitle{Gradient-based learning applied to document
  recognition}.
\newblock \bibinfo{journal}{\emph{Proc. IEEE}} \bibinfo{volume}{86},
  \bibinfo{number}{11} (\bibinfo{year}{1998}), \bibinfo{pages}{2278--2324}.
\newblock


\bibitem[\protect\citeauthoryear{Liang, Wang, Chen, Jiang, and Zhang}{Liang
  et~al\mbox{.}}{2018}]%
        {liang2018fuzz}
\bibfield{author}{\bibinfo{person}{Jie Liang}, \bibinfo{person}{Mingzhe Wang},
  \bibinfo{person}{Yuanliang Chen}, \bibinfo{person}{Yu Jiang}, {and}
  \bibinfo{person}{Renwei Zhang}.} \bibinfo{year}{2018}\natexlab{}.
\newblock \showarticletitle{Fuzz testing in practice: Obstacles and solutions}.
  In \bibinfo{booktitle}{\emph{2018 IEEE 25th International Conference on
  Software Analysis, Evolution and Reengineering (SANER)}}. IEEE,
  \bibinfo{pages}{562--566}.
\newblock


\bibitem[\protect\citeauthoryear{Mnih, Kavukcuoglu, Silver, Rusu, Veness,
  Bellemare, Graves, Riedmiller, Fidjeland, Ostrovski, et~al\mbox{.}}{Mnih
  et~al\mbox{.}}{2015}]%
        {mnih2015human}
\bibfield{author}{\bibinfo{person}{Volodymyr Mnih}, \bibinfo{person}{Koray
  Kavukcuoglu}, \bibinfo{person}{David Silver}, \bibinfo{person}{Andrei~A
  Rusu}, \bibinfo{person}{Joel Veness}, \bibinfo{person}{Marc~G Bellemare},
  \bibinfo{person}{Alex Graves}, \bibinfo{person}{Martin Riedmiller},
  \bibinfo{person}{Andreas~K Fidjeland}, \bibinfo{person}{Georg Ostrovski},
  {et~al\mbox{.}}} \bibinfo{year}{2015}\natexlab{}.
\newblock \showarticletitle{Human-level control through deep reinforcement
  learning}.
\newblock \bibinfo{journal}{\emph{Nature}} \bibinfo{volume}{518},
  \bibinfo{number}{7540} (\bibinfo{year}{2015}), \bibinfo{pages}{529}.
\newblock


\bibitem[\protect\citeauthoryear{Moosavi~Dezfooli, Fawzi, and
  Frossard}{Moosavi~Dezfooli et~al\mbox{.}}{2016}]%
        {moosavi2016deepfool}
\bibfield{author}{\bibinfo{person}{Seyed~Mohsen Moosavi~Dezfooli},
  \bibinfo{person}{Alhussein Fawzi}, {and} \bibinfo{person}{Pascal Frossard}.}
  \bibinfo{year}{2016}\natexlab{}.
\newblock \showarticletitle{Deepfool: a simple and accurate method to fool deep
  neural networks}. In \bibinfo{booktitle}{\emph{Proceedings of 2016 IEEE
  Conference on Computer Vision and Pattern Recognition (CVPR)}}.
\newblock


\bibitem[\protect\citeauthoryear{Pei, Cao, Yang, and Jana}{Pei
  et~al\mbox{.}}{2017}]%
        {pei2017deepxplore}
\bibfield{author}{\bibinfo{person}{Kexin Pei}, \bibinfo{person}{Yinzhi Cao},
  \bibinfo{person}{Junfeng Yang}, {and} \bibinfo{person}{Suman Jana}.}
  \bibinfo{year}{2017}\natexlab{}.
\newblock \showarticletitle{Deepxplore: Automated whitebox testing of deep
  learning systems}. In \bibinfo{booktitle}{\emph{Proceedings of the 26th
  Symposium on Operating Systems Principles}}. ACM, \bibinfo{pages}{1--18}.
\newblock


\bibitem[\protect\citeauthoryear{Simonyan and Zisserman}{Simonyan and
  Zisserman}{2015}]%
        {simonyan2014very}
\bibfield{author}{\bibinfo{person}{Karen Simonyan} {and}
  \bibinfo{person}{Andrew Zisserman}.} \bibinfo{year}{2015}\natexlab{}.
\newblock \showarticletitle{Very deep convolutional networks for large-scale
  image recognition}. In \bibinfo{booktitle}{\emph{Proceedings of the
  International Conference on Learning Representations}}.
\newblock


\bibitem[\protect\citeauthoryear{Sutskever, Vinyals, and Le}{Sutskever
  et~al\mbox{.}}{2014}]%
        {sutskever2014sequence}
\bibfield{author}{\bibinfo{person}{Ilya Sutskever}, \bibinfo{person}{Oriol
  Vinyals}, {and} \bibinfo{person}{Quoc~V Le}.}
  \bibinfo{year}{2014}\natexlab{}.
\newblock \showarticletitle{Sequence to sequence learning with neural
  networks}. In \bibinfo{booktitle}{\emph{Advances in neural information
  processing systems}}. \bibinfo{pages}{3104--3112}.
\newblock


\bibitem[\protect\citeauthoryear{Szegedy, Zaremba, Sutskever, Bruna, Erhan,
  Goodfellow, and Fergus}{Szegedy et~al\mbox{.}}{2013}]%
        {szegedy2013intriguing}
\bibfield{author}{\bibinfo{person}{Christian Szegedy},
  \bibinfo{person}{Wojciech Zaremba}, \bibinfo{person}{Ilya Sutskever},
  \bibinfo{person}{Joan Bruna}, \bibinfo{person}{Dumitru Erhan},
  \bibinfo{person}{Ian Goodfellow}, {and} \bibinfo{person}{Rob Fergus}.}
  \bibinfo{year}{2013}\natexlab{}.
\newblock \showarticletitle{Intriguing properties of neural networks}.
\newblock \bibinfo{journal}{\emph{arXiv preprint arXiv:1312.6199}}
  (\bibinfo{year}{2013}).
\newblock


\bibitem[\protect\citeauthoryear{Tian, Pei, Jana, and Ray}{Tian
  et~al\mbox{.}}{2018}]%
        {tian2018deeptest}
\bibfield{author}{\bibinfo{person}{Yuchi Tian}, \bibinfo{person}{Kexin Pei},
  \bibinfo{person}{Suman Jana}, {and} \bibinfo{person}{Baishakhi Ray}.}
  \bibinfo{year}{2018}\natexlab{}.
\newblock \showarticletitle{Deeptest: Automated testing of
  deep-neural-network-driven autonomous cars}. In
  \bibinfo{booktitle}{\emph{Proceedings of the 40th International Conference on
  Software Engineering}}. ACM, \bibinfo{pages}{303--314}.
\newblock


\bibitem[\protect\citeauthoryear{Wang, Liang, Chen, Jiang, Jiao, Liu, Zhao, and
  Sun}{Wang et~al\mbox{.}}{2018}]%
        {wang2018safl}
\bibfield{author}{\bibinfo{person}{Mingzhe Wang}, \bibinfo{person}{Jie Liang},
  \bibinfo{person}{Yuanliang Chen}, \bibinfo{person}{Yu Jiang},
  \bibinfo{person}{Xun Jiao}, \bibinfo{person}{Han Liu}, \bibinfo{person}{Xibin
  Zhao}, {and} \bibinfo{person}{Jiaguang Sun}.}
  \bibinfo{year}{2018}\natexlab{}.
\newblock \showarticletitle{SAFL: increasing and accelerating testing coverage
  with symbolic execution and guided fuzzing}. In
  \bibinfo{booktitle}{\emph{Proceedings of the 40th International Conference on
  Software Engineering: Companion Proceeedings}}. ACM, \bibinfo{pages}{61--64}.
\newblock


\bibitem[\protect\citeauthoryear{Wicker, Huang, and Kwiatkowska}{Wicker
  et~al\mbox{.}}{2018}]%
        {wicker2018feature}
\bibfield{author}{\bibinfo{person}{Matthew Wicker}, \bibinfo{person}{Xiaowei
  Huang}, {and} \bibinfo{person}{Marta Kwiatkowska}.}
  \bibinfo{year}{2018}\natexlab{}.
\newblock \showarticletitle{Feature-guided black-box safety testing of deep
  neural networks}. In \bibinfo{booktitle}{\emph{International Conference on
  Tools and Algorithms for the Construction and Analysis of Systems}}.
  Springer, \bibinfo{pages}{408--426}.
\newblock


\bibitem[\protect\citeauthoryear{Yuan, Lu, Wang, and Xue}{Yuan
  et~al\mbox{.}}{2014}]%
        {yuan2014droid}
\bibfield{author}{\bibinfo{person}{Zhenlong Yuan}, \bibinfo{person}{Yongqiang
  Lu}, \bibinfo{person}{Zhaoguo Wang}, {and} \bibinfo{person}{Yibo Xue}.}
  \bibinfo{year}{2014}\natexlab{}.
\newblock \showarticletitle{Droid-sec: deep learning in android malware
  detection}. In \bibinfo{booktitle}{\emph{ACM SIGCOMM Computer Communication
  Review}}, Vol.~\bibinfo{volume}{44}. ACM, \bibinfo{pages}{371--372}.
\newblock


\bibitem[\protect\citeauthoryear{Zalewski}{Zalewski}{2007}]%
        {zalewski2007american}
\bibfield{author}{\bibinfo{person}{Michal Zalewski}.}
  \bibinfo{year}{2007}\natexlab{}.
\newblock \bibinfo{title}{American fuzzy lop}.
\newblock
\newblock


\end{thebibliography}

\end{document}